\documentclass[12pt,a4paper]{iopart}

\usepackage{amssymb}
\usepackage{bm}
\usepackage{bbold}
\usepackage{graphicx}
\usepackage{braket}
\usepackage[colorlinks, linkcolor=blue, citecolor=blue, urlcolor=blue, breaklinks=true]{hyperref}

\begin{document}

\title[Dynamics of correlations in 2D: a phase-space Monte-Carlo study]{Dynamics of correlations in two-dimensional quantum spin models with long-range interactions:
A phase-space Monte-Carlo study}

\author{J.~Schachenmayer, A.~Pikovski, and A.~M.~Rey}

\address{JILA, NIST \& Department of Physics, University of Colorado, 440 UCB, Boulder, CO 80309, USA}
\ead{johannes.schachenmayer@gmail.com}
\date{\today}

\begin{abstract}
Interacting quantum spin models  are remarkably useful for describing  different types of physical, chemical, and biological  systems. 
Significant understanding of  their equilibrium properties  has been achieved to date,  especially for the case of  spin models with short-range couplings. However, progress towards  the development of a comparable understanding  in long-range interacting models, in particular out-of-equilibrium, remains limited.  In a recent work, we proposed a semiclassical numerical method to study  spin models, the discrete truncated Wigner approximation (DTWA),  and demonstrated its  capability to correctly capture the dynamics of  one- and two-point correlations  in one dimensional (1D) systems. Here we go one step forward and use the DTWA method to study the dynamics of correlations in 2D systems with many spins and different types of long-range couplings, in regimes where other numerical methods  are  generally unreliable.  We compute spatial and time-dependent correlations for spin-couplings that decay  with distance as a power-law and  determine  the velocity at which  correlations propagate through
the system.  Sharp changes in the behavior of those  velocities are found as a function of the power-law decay  exponent.  Our predictions are relevant for a broad range of systems including  solid state materials,   atom-photon systems and  ultracold gases of  polar molecules, trapped ions, Rydberg, and  magnetic atoms. We validate the DTWA predictions for small 2D systems and  1D   systems, but ultimately, in the spirt of quantum simulation,  experiments will be   needed to confirm our predictions for large 2D systems.
\end{abstract}


\maketitle

\section{Introduction}

An important   advance towards  understanding  non-equilibrium phenomena  has been made possible by  recent advances in cooling,  trapping and manipulating atomic, molecular and optical (AMO) systems \cite{Bloch2008r}.  In contrast to solid state systems, where studying equilibrium situations is the default approach (given their complex environment and fast relaxation rates),   AMO systems  provide a unique platform to observe and investigate
non-equilibrium quantum dynamics in strongly interacting
many-body systems~\cite{lamacraft-moore:potential-insights_2012}. 
Their high dynamic tunability even during the course of an experiment, their decoupling from the external environment and their characteristic low-energy scales, lead to long non-equilibrium time-scales over which the system can be followed almost in real time.
Quantum quenches,
i.e.~the dynamics induced by abruptly changing parameters of the system,
is currently a common  protocol used  to probe AMO systems.

One of the most promising opportunities offered by modern  AMO physics is the ability to engineer   interatomic interactions different from the standard contact and isotropic interactions arising from ultracold collisions. At the heart of  this capability are recent experimental developments on controlling AMO systems with complex internal structure and with   enlarged sets of  degrees of freedom such as  polar molecules~\cite{yan:observation_2013}, trapped ions~\cite{kim:quantum_2010,britton:engineered_2012,lanyon:universal_ion_2011}, magnetic atoms~\cite{lu:strongly_2011,aikawa:bose-einstein_2012,depaz2013}, Rydberg atoms~\cite{schwarzkopf:imaging_2011,schwarzkopf:spatial_2013,Schausz:obervation_2012,anderson:dephasing_2002,
butscher:atom-molecule_2010,nipper:highly_2012},  and  alkaline earth atoms~\cite{swallows:suppression_2011,lemke:p-wave_2011,martin:quantum_2013,Rey_arXiv_2013,Olmos2013}. All these systems have  in common that they can exhibit long-range interactions. This experimental progress is opening new frontiers, and at the same time demanding for improved theoretical techniques, that are capable of dealing  with the complicated non-equilibrium quantum dynamics of  long-range interacting systems.

In a prior work \cite{schachenmayer2015}, we proposed a semiclassical phase-space method to study non-equilibrium quantum dynamics. We used this numerical method, that we named the discrete truncated Wigner approximation (DTWA),  to study the dynamics of  single particle observables and correlation functions after a quench. The DTWA was  benchmarked in one-dimensional spin models with numerically exact time-dependent density matrix renormalization group calculations (t--DMRG)~\cite{vidal2004, white:real-time_2004,daley:time-dependent_2004,schollwoeck:density-matrix_2011}  and  excellent  agreement was found.

In this work, we generalize the calculations to two-dimensional (2D) Ising and  XY  spin models with various ranges of interactions. We study the time-evolution in a setup that is equivalent to  a Ramsey-type  procedure, as realized in recent experiments. This dynamical protocol has been   used, for example,  to observe dipolar spin-exchange interactions in ultracold molecules~\cite{yan:observation_2013,Hazzard2013,hazzard:many-body_2014}, to benchmark the Ising dynamics  of hundreds of trapped ions~\cite{britton:engineered_2012}, and to precisely measure atomic transitions as well as many-body interactions  in optical lattice clocks~\cite{hinkley:atomic_2013,bloom:optical_2014,martin:quantum_2013,lemke:p-wave_2011}. Using the DTWA  we compute the dynamics of the collective spin as well as spatially resolved  two-point
correlation functions. Since  the applicability of the  t-DMRG method  becomes limited in 2D, to benchmark the DTWA we  perform  numerical comparisons with small systems (where exact diagonalization is possible), and with the analytically solvable Ising case \cite{Lowe-Norberg-1957,FossFeig_NJP2013,vdW_NJP2013}. We then extend the calculations to large XY spin models, and find remarkably sharp changes in the propagation of correlations  as we vary the power law-decay exponent of the interactions.

 The dynamics of the two-point correlations is directly linked to  the speed of  propagation of information in quantum many-body systems, a topic of great interest to quantum information science and currently  subjected to intensive investigation.
For systems  with short-range interactions, there is a well understood bound (derived by Lieb and Robinson) that limits correlations to remain within a linear effective ``light cone'' region \cite{HastingsKoma_2006,Hastings2010}. On the  contrary there are many open questions
about what limits  the propagation of information in quantum many-body systems with long-range interactions \cite{eisert_breakdown_2013,schachenmayer_entanglement_2013,Hauke2013,Foss-Feig2014,gong_persistence_2014,Richerme2014,Jurcevic2014,carleo_light-cone_2014,Garcia2013,storch_interplay_2015,cevolani_protected_2015}. Here we use the DTWA method to determine the shape of the causal region
and the speed at which correlations propagate after a global quench. We compute the full crossover of the dynamics when changing the range of the interactions over a large range in 1D and 2D systems, and observe remarkable agreement of DTWA results with t-DMRG predictions in the 1D case. Our calculations and their natural variations (e.g.~local instead of global quenches)  should be testable in experiments with polar molecules and trapped ions in the near future.

This paper is organized as follows: In section~\ref{sec:model} we introduce the models that are studied. In section~\ref{sec:twa}  we review the  DTWA technique that was  introduced in Ref.~\cite{schachenmayer2015}.  We benchmark the DTWA method by comparing the dynamics of single-spin observables and  correlation functions  with exact solutions in section~\ref{sec:bench}.  In  section~\ref{sec:large} those calculations are extended to large systems where currently no other method is applicable. In section~\ref{sec:crossover} we use the DTWA for a systematic calculation of the  light-cone dynamics as we vary  the range of the interactions.
Finally, section~\ref{sec:concl} concludes and provides an outlook.

\section{Spin models and dynamics}
\label{sec:model}

We will focus our attention on Hamiltonians that fall under the generic heading of  spin-$1/2$ XXZ models given by ($\hbar=1$)
\begin{eqnarray}\label{H}
\hat{H}=\frac{1}{2} \sum_{i \neq j} \left[ {J^\perp_{ij}} ( \hat{\sigma}_i^x \hat{\sigma}_j^x + \hat{\sigma}_i^y \hat{\sigma}_j^y)   +  J^z_{ij} \, \hat{\sigma}_i^z \hat{\sigma}_j^z  \right],
\end{eqnarray} where the sum extends over all pairs of sites of an arbitrary lattice, $ \hat{\sigma}_i^{x,y,z}$ are Pauli matrices for the spin on site $i$,
 $J_{ij}=J_{ji}$ and $J_{ii}=0$.
In our analysis the interactions are assumed to decay as a function of the distance with a decay exponent $\alpha$, that is $J_{ij}^{\perp,z}\equiv J (a/|{\bf r}_{ij}|)^{\alpha}$. Here, ${\bf r}_{ij}$ is the vector connecting spins on sites  $i$ and $j$ and $a$ is the lattice spacing. We  concentrate  our study on two specific  cases, Ising ($J^\perp_{ij}=0$) and XY ($J^z_{ij}=0$) interactions. We consider a general 2D grid, i.e.~a lattice with $N_x\times N_y=M$ sites with spins at positions ${\bf r}_i=a(n_x,n_y)$ where $n_{x,y}$ are  integers. The lattice spacing $a$ is set to $1$ throughout this paper.

Spin-$1/2$ XXZ models broadly describe a variety of physical systems. For instance, in the AMO context, XXZ spin Hamiltonians have been used to model the dynamics of  ultracold molecules in optical lattices, trapped ions, Rydberg atoms, neutral atoms in optical clocks, and ultracold magnetic atoms. A summary of how these models  are realized in those systems can be found in Ref.~\cite{Hazzard2014b}. Here we describe  the two most relevant ones for this work: ultracold polar molecules and trapped ions.

 In {\it ultracold polar molecules} pinned  in optical lattices, the spin-1/2 degree of freedom  can be encoded in  two rotational states, and the spin--spin couplings are generated by dipolar interactions. The difference in dipole moments between the two states (which arises in the presence of an electric field) generates the Ising term while transition dipole moments between the two rotational  states (which can exist even in the absence of an electric field)   give rise to the spin-exchange terms~\cite{barnett:quantum_2006}. The ratio between the Ising and XY couplings can be manipulated using electromagnetic fields \cite{wall:hyperfine_2010,Schachenmayer:dynamical_2010,perez-rios:external_2010,gorshkov:quantum_2011,gorshkov:tunable_2011,Review2014}.  The case without electric field which implements the pure  XY model has been recently  realized  with  KRb  polar  molecules in a 3D optical lattice \cite{yan:observation_2013,hazzard:many-body_2014}. In general, dipolar interactions are long-ranged and spatially anisotropic. In a 2D geometry, however, they become isotropic if the electric field that sets the quantization axis is set  perpendicular to the plane containing the molecules. This is the case considered throughout this paper.

{\em Crystals of 2D self-assembled trapped ions}  can also be used to implement specific cases of equation~(\ref{H}), cf.~\cite{britton:engineered_2012}. By addressing the ions confined by  a Penning trap with a spin-dependent optical potential, the vibrations of the crystal mediate a long-range Ising interaction that can be approximately described by a power-law with $0\le \alpha < 3$ ~\cite{porras:quantum_2006,kim:entanglement_2009,kim:quantum_2010,barreiro:open-system_2011,britton:engineered_2012,islam:emergence_2013}. To engineer an XY model, one needs  to add a strong transverse  field that projects out the off-resonant terms  in the Ising interactions that change the  magnetization
along the field quantization direction \cite{Richerme2014,Jurcevic2014}.

The dynamical procedure considered here, is identical to a Ramsey spectroscopy setup. It has been implemented in  various recent  experiments as a diagnostic tool for interactions~\cite{hazzard:many-body_2014}. It consists of preparing an initial state with
all spins  aligned  (at time $t=0$) along a specific direction, here we  consider it to be the $x$ direction, i.e.~$\ket{\psi(t=0)}=\bigotimes_i^M(\ket{\uparrow}_i+\ket{\downarrow}_i)/\sqrt{2}$. Then this
initial state evolves under the Ising or XY Hamiltonian (\ref{H}) for a time $t$, leading to the state $\ket{\psi(t)}=\exp(-{\rm i}t \hat H) \ket{\psi(t=0)}$. Afterwards one measures expectation values of an observable with the time-evolved state, $\bra{\psi(t)} \hat{O} \ket{\psi(t)} $. In this paper, we focus on two-point correlations and the collective spin along $x$ as observables.

\section{The DTWA method}
\label{sec:twa}

Phase space methods, such as the truncated Wigner approximation, solve the quantum  dynamics approximately by replacing the time-evolution by a semi-classical evolution via classical trajectories. The quantum uncertainty in the initial state is accounted for by an average over different initial conditions~\cite{blakie_dynamics_2008,polkovnikov:phase_2010}, determined by the Wigner function. Although the truncated Wigner approximation was initially developed to deal with systems with continuous degrees of freedom~\cite{graham1973,steel1998}, it has also been adopted to treat collective spin models. In this case  the standard method has been to approximate the  Wigner function by a continuous Gaussian distribution that facilitates the sampling of trajectories \cite{polkovnikov:phase_2010}.
This continuous approximation, however,  misses important aspects inherent to the discrete nature of spin variables and is unsuitable for systems with finite-range interactions. To deal with more generic types of spin models,  recently we proposed instead to sample a discrete Wigner function for each spin and named this approach the DTWA method \cite{schachenmayer2015}.  In this section we  present an overview of  the DTWA method. For details the reader is referred to Ref.~\cite{schachenmayer2015}.

Operators and wave functions on the Hilbert space of a quantum system can be, equivalently,
represented (mapped) on a classical  phase space. There, any operator $\hat{O}$ corresponds to a real-valued function of the classical phase-space variables, ${\mathcal O}^W$ (a so-called Weyl symbol). The phase-space function corresponding to the density matrix is precisely the Wigner function $w$.
For continuous variables $p,q$ in one dimension (for simplicity of presentation), the expectation value of an operator can be exactly represented as
$
\langle \hat{{O}}\rangle(t) = \int \int \! dp dq  \, w(p,q;t) \mathcal{O}^W(p,q)
$.

For quantum systems with discrete degrees of freedom, one can introduce a ``discrete phase-space'' in various ways, see~\cite{wootters_picturing_2003} and references therein.
Here we use the representation of  Wootters~\cite{wootters_wigner-function_1987,wootters_picturing_2003}. For a single spin-1/2, it uses four distinct phase-points, and
all phase-space functions are thus defined as $2\!\times\!2$ matrices. In our approximation, the phase-space of $N$ spins factorizes into a product of $N$ phase-spaces for each individual spin.

In both continuous and discrete variables  however, it is not possible to compute the time-evolution exactly. The spirit of the truncated Wigner approximation and the DTWA  is  to  take quantum fluctuations into account only to lowest order \cite{polkovnikov:phase_2010}. In particular, we switch to a ``Heisenberg picture'', such that the Wigner function does not evolve in time (i.e.\ it is fixed to the initial state) while the operator-functions are time-dependent. The approximation that we make is to assume that the operators in phase space follow their  classical evolution:
\begin{equation}
	\langle \hat O \rangle (t) =\sum_{ \gamma} w(\gamma;0) \mathcal{O}^W(\gamma;t)
	\approx \sum_{ \gamma} w(\gamma;0) \mathcal{O}^{W,{\rm cl.}}(\gamma;t),
	\label{eq:dtwa}
\end{equation}
where $\gamma$ runs over the points of the discrete phase space, $w(\gamma;0)$ is the Wigner function at $t=0$ on the  discrete many-body phase space, and $\mathcal{O}^{W,{\rm cl.}}(\gamma;t)$ is the classically evolved operator-function (Weyl symbol) that corresponds to our observable.

Equation~(\ref{eq:dtwa}) is solved numerically by choosing a large number $n_t$ of random initial spin-configurations, with probability according to  $w(\gamma;0)$. Each of this ``Monte-Carlo trajectories'' is evolved independently following the classical equations of motion (see below). The expectation value in equation~(\ref{eq:dtwa}) is calculated by averaging. We find that the number of required trajectories, $n_t$, does not depend on the system size, but rather on the observable under consideration.

To apply the truncated Wigner approximation we have to compute the classical equations of motion for the spin components of each spin $i$: $s^x_i,s^y_i, s^z_i$.
One way to do this is to replace spin operators $\vec{\sigma}$ by classical variables ($\vec{s}=\langle \vec{\sigma} \rangle$) in the Hamiltonian and to compute the Poisson bracket~\cite{polkovnikov:phase_2010}, giving
$
\dot{s}^\delta_i=
2 \sum_\beta \epsilon_{\delta \beta \gamma} s^\gamma_i\frac{\partial H}{\partial s^\beta_i}
$
with $\epsilon$  the fully antisymmetric tensor. Alternatively, the same classical (mean-field) equations of motion can be obtained
 from a  product state ansatz of the  density matrix \cite{schachenmayer2015}.
For the Ising interaction Hamiltonian the classical equations for the spin components are given by:
\begin{eqnarray}
	\dot s^x_n &=   -  2s^y_n  \sum_ m J^z_{n,m} s^z_m \equiv -2s^y_n \beta^z_n,  \\
	\dot s^y_n &=  2 s^x_n  \sum_ m J^z_{n,m} s^z_m \equiv 2s^x_n \beta^z_n, \\
	\dot s^z_n &= 0,
\end{eqnarray} where we introduced the quantity  $\beta^{\delta=x,y,z}_n\equiv \sum_ m J^z_{n,m} s^{\delta=x,y,z}_m$ which can be interpreted as an effective magnetic mean-field acting on spin $n$ induced by the other spins. For the XY interaction the classical equations of motion are given by:
\begin{eqnarray}
	\dot s^x_n &=   2 s^z_n \sum_ m J^\perp_{n,m} s^y_m \equiv 2 s^z_n \beta^y_n, \label{eq:mf1} \\
	\dot s^y_n &=   - 2 s^z_n  \sum_ m J^\perp_{n,m} s^x_m \equiv - 2 s^z_n \beta^x_n, \label{eq:mf2}\\
	\dot s^z_n &= 2 \sum_ m J^\perp_{n,m} (s^x_m s^y_n - s^y_m s^x_n) \equiv 2 s^y_n  \beta^x_n -  2s^x_n \beta^y_n \label{eq:mf3}
	.
\end{eqnarray}
Note that the sums exclude the term $m=n$ since we set $J_{nn}=0$.

In practice, applying the DTWA means to solve these equations of motion $n_t$ times, while each time choosing a different random initial configuration. For our particular initial state (pointing along the $x$ direction), the prescription for the correct sampling is to randomly pick values of the orthogonal spin-components for each spin from $s^y_n(0),s^z_n(0)\in\{-1,+1\}$.

The error of the DTWA method, i.e.\ the deviation from the exact solution arises entirely due to the semiclassical approximation for the time-evolution [cf.\ Eq.~(\ref{eq:dtwa})]. The Wigner function of the initial state, on the other hand, is sampled exactly here up to statistical errors (which can be controlled by increrasing the number of trajectories). For the exactly solvable Ising model, the DTWA method turns out to be able to reproduce the exact solution for single-particle observables; for two-particle correlations the error can be given explicitly  \cite{schachenmayer2015} (see below).

We note that the  DTWA  clearly goes beyond the mean-field predictions. A pure mean-field theory  is not only incapable to capture spin-spin correlations (they are all zero due to the factorization approximation of the density matrix)  but even the single particle mean-field obsevables can be completely incorrect.  For example equations (\ref{eq:mf1}--\ref{eq:mf3})] predict no dynamics at all in our Ramsey setup where the collective Bloch vector points initially along $x$.

\section{Benchmarking the DTWA}
\label{sec:bench}

\subsection{Contrast}

\begin{figure}[tb]
  \centering
  \includegraphics[width=0.7\textwidth]{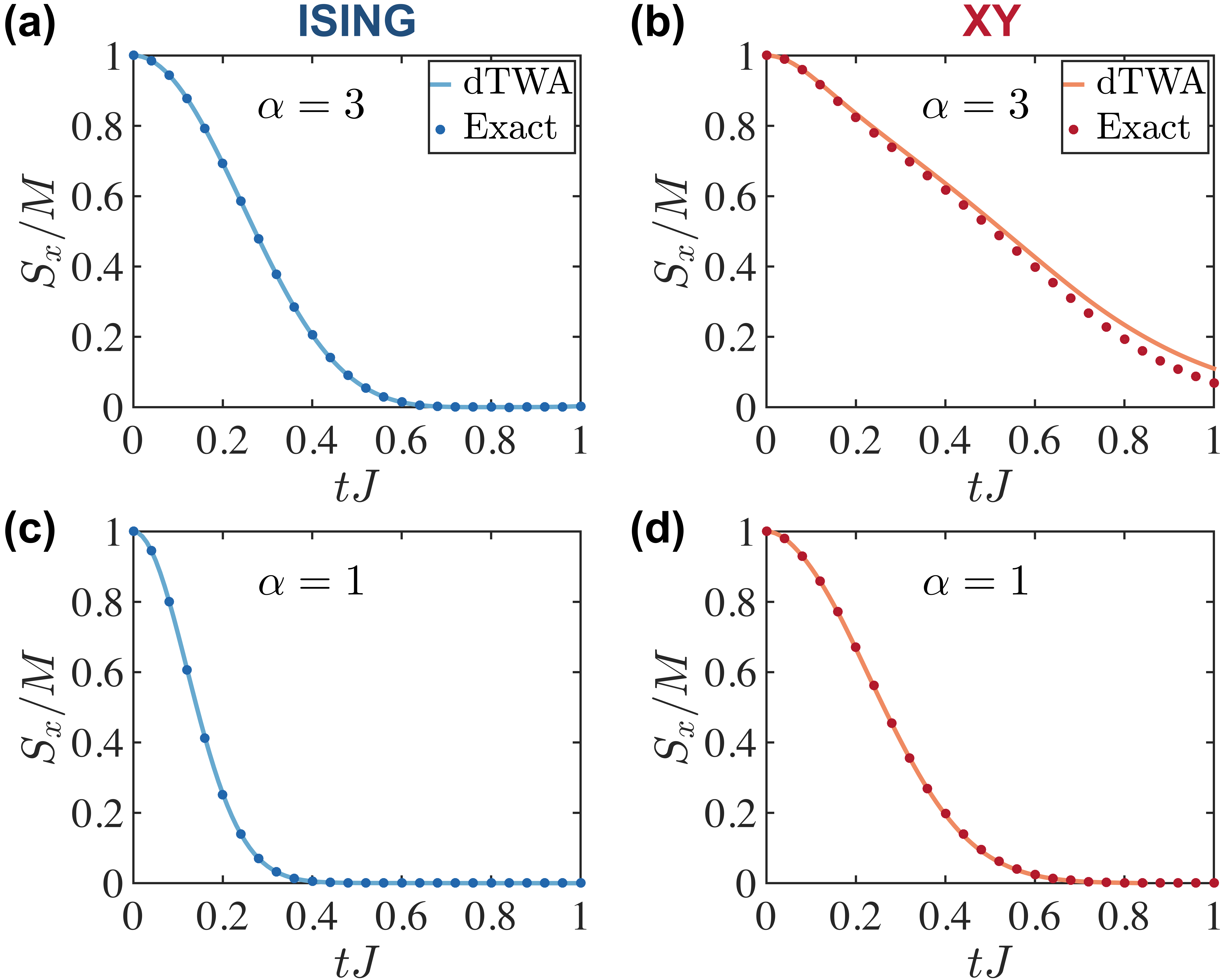}
  \caption{Time-evolution of the total spin in the $x$-direction (or Ramsey contrast), $S_x=\langle \hat S_x \rangle$. Compared are exact solutions (points) and DTWA results (lines). (a,c) Ising interactions on a $31\times31$ lattice with $\alpha=3$ (panel a) and $\alpha=1$ (panel c). (b,d) XY interactions on a $4\times 5$ lattice with $\alpha=3$ (panel b) and $\alpha=1$ (panel d). \label{fig:benchmark_sx_xy_ising}}
\end{figure}

Before discussing results obtained by the DTWA method for  the propagation of correlations through a 2D system,
we need to consider how well this approximate method performs in such a setting. As a starting point we first consider  simpler single particle  observables. One  observable with immediate relevance to AMO experiments is the ``contrast'' or amplitude of the oscillations in a Ramsey experiment, which for our initial state is  given by $\hat S_x= \sum_i \hat \sigma^x_i$. In figure~\ref{fig:benchmark_sx_xy_ising} we compare the time-evolution of $\langle \hat S_x \rangle$ to exact solutions. For Ising interactions, $J^\perp_{ij}=0$ in equation~(\ref{H}),  exact analytical expressions for the dynamics exists~\cite{Lowe-Norberg-1957,FossFeig_PRA2013,FossFeig_NJP2013}. We can thus compare our DTWA results in a large $31\times 31$ system. In contrast, for XY interactions [$J^z_{ij}=0$ in equation~(\ref{H})],  no analytical solution is known in 2D. Therefore, in this case we  resort to comparisons with a numerically exact diagonalization (ED) methods which is limited to small systems sizes. In this case, we choose a $4\times5$ lattice.

To cover different regimes, in figure~\ref{fig:benchmark_sx_xy_ising} we consider the  Ising (panels a,c) and  the XY (panels b,d) cases with two different power-law decay exponent, $\alpha=3$ (panels a, b), and $\alpha=1$  (panels c, d). Remarkably, the Ising dynamics is exactly covered by the DTWA approximation, an agreement that  can be rigorously  justified \cite{schachenmayer2015}. For the XY case we also find excellent agreement. While for $\alpha=3$ small numerical differences are visible, for $\alpha=1$, the different curves are nearly indistinguishable.

\subsection{Spatial correlations}

\begin{figure}[tb]
  \centering
    \includegraphics[width=0.7\textwidth]{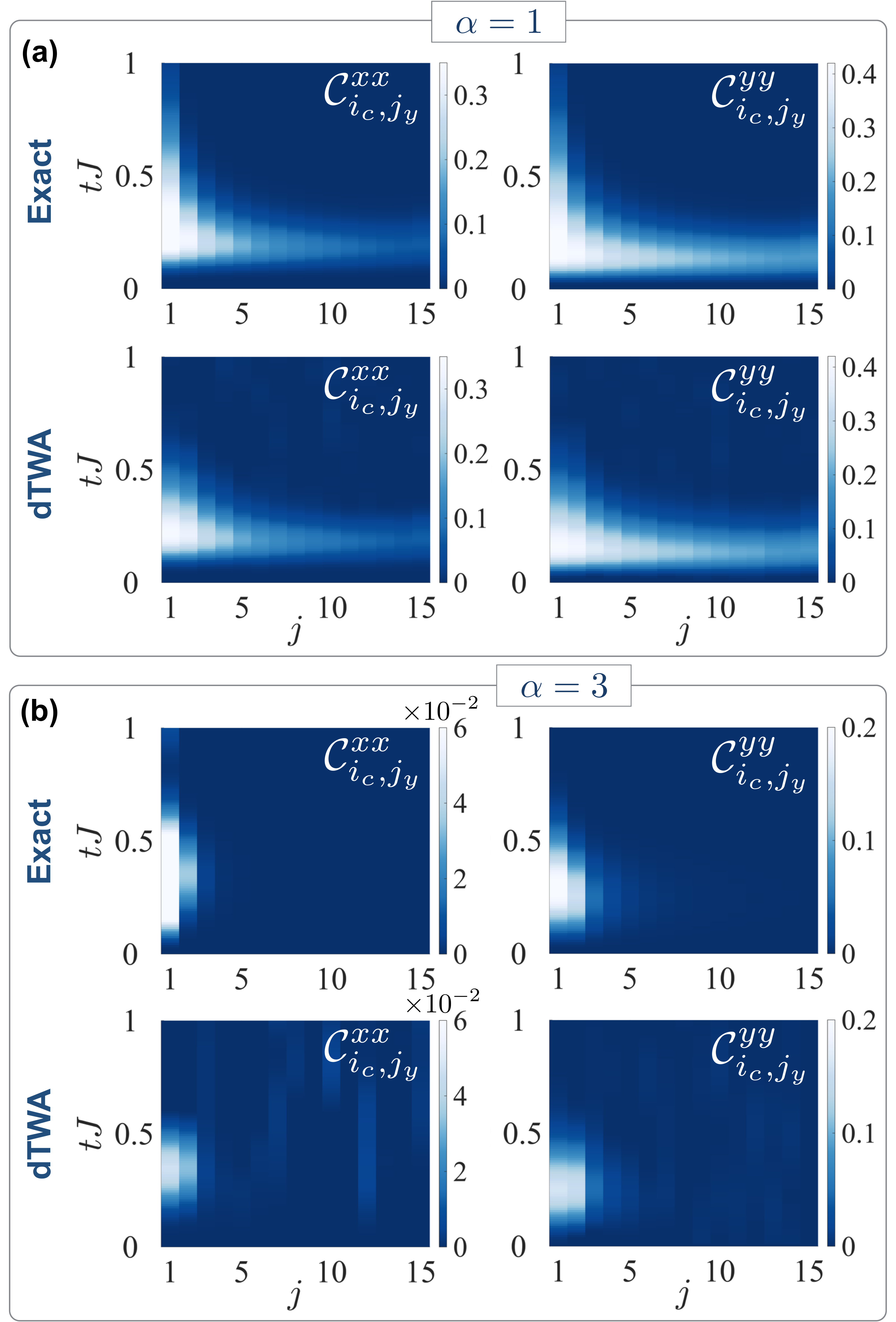}
  \caption{Ising interaction benchmark: Time-evolution of connected correlation functions $\mathcal{C}^{xx,yy}_{i_c,j_y}$ in a $31\times31$ lattice between the center site ${\bf r}_{i_c}=(16,16)$ and the sites ${\bf r}_{j_y}=(16,16+j)$.  Compared are exact solutions (upper plots in each panel) and DTWA results (lower plots). (a) $\alpha=1$, (b) $\alpha=3$. The left and right column depicts $\mathcal{C}^{xx}_{i_c,j_y}$ and $\mathcal{C}^{yy}_{i_c,j_y}$, respectively. \label{fig:benchmark_corr_ising_al3al1}}
\end{figure}

We now turn to checking the capability  of the DTWA to describe the time evolution of spatial two-point correlations. We again first consider  the case of Ising interactions where we can compare the DTWA to an exact analytical solution in large systems~\cite{vdW_NJP2013,FossFeig_NJP2013}. In particular we consider a $31\times 31$  square lattice geometry. We study the  time-dependence of  connected correlation functions between sites $n$ and $m$;
\begin{eqnarray}
	\mathcal{C}^{\beta \beta}_{n,m}\equiv \langle \hat \sigma_n^{\beta}  \hat \sigma_m^{\beta} \rangle -  \langle \hat \sigma_n^{\beta} \rangle \langle  \hat \sigma_m^{\beta} \rangle \qquad \beta\in\{x,y,z\},
\end{eqnarray} we calculate the correlations from the central spin of the system,  ${\bf r}_{n}=(16,16)$, along the $y$-direction, ${\bf r}_{m}=(16,16+j)$ (note that results along the $x$-direction are identical).

The comparisons for Ising interactions are summarized in figure~\ref{fig:benchmark_corr_ising_al3al1}. We find that in particular for substantially long-range interactions (case $\alpha=1$) the DTWA gives excellent results. Only for $j\sim1,2$ there are small quantitative differences (see below). For shorter range interactions (case $\alpha=3$) the correlations between long-distance spins remain essentially zero. The  relevant  short-distance correlations are reproduced  by the   DTWA with small deviations that are comparable to  those ones seen  in the  long-range interacting cases. Overall, even for  $\alpha=3$, the spreading of the correlations is  very well (qualitatively) reproduced by the DTWA. Note that the fluctuations around zero for the DTWA solution are statistical because of the finite number of trajectories, which is of order $\mathcal{O}(10^5)$ in all our calculations.

 By comparing the exact analytical solution of the problem with the DTWA prediction one finds \cite{schachenmayer2015} that
$
\langle \hat \sigma_i^\pm \hat \sigma_j^\pm \rangle (t)_{\rm DTWA} = \langle \hat \sigma_i^\pm \hat \sigma_j^\pm \rangle (t)_{\rm exact} \cos^2(2tJ^z_{ij}).
$
For our particular problem (note that $\langle\hat \sigma_i^y \rangle=0$ for all times), this implies that the relative error in the $\mathcal{C}^{yy}_{i,j}$  correlations can be quantified as
\begin{eqnarray}
\label{eq:err}
\epsilon_j^{yy} \equiv \left| \frac{\mathcal{C}^{yy, {\rm exact}}_{i_c,j_y} - \mathcal{C}^{yy, {\rm DTWA}}_{i_c,j_y} }{\mathcal{C}^{yy, {\rm exact}}_{i_c,j_y}}\right|= |1-\cos^2(2tJ^z_{i_cj_y})|.
\end{eqnarray}
Since $J^z_{ij}$  decays as a power-law  with the distance between spins $i$ and $j$,  for short times, $t\lesssim J^{-1}$, the relative error decreases with distance as well. In particular for $tJ\ll1$ it follows that $\epsilon_j^{yy}\propto (tJ)^2 j^{-2\alpha}$. Note that equation~(\ref{eq:err}) also implies that $\epsilon_j^{yy}$  only depends on the coupling-strength between the two spins in consideration.

\begin{figure}[tb]
  \centering
    \includegraphics[width=0.7\textwidth]{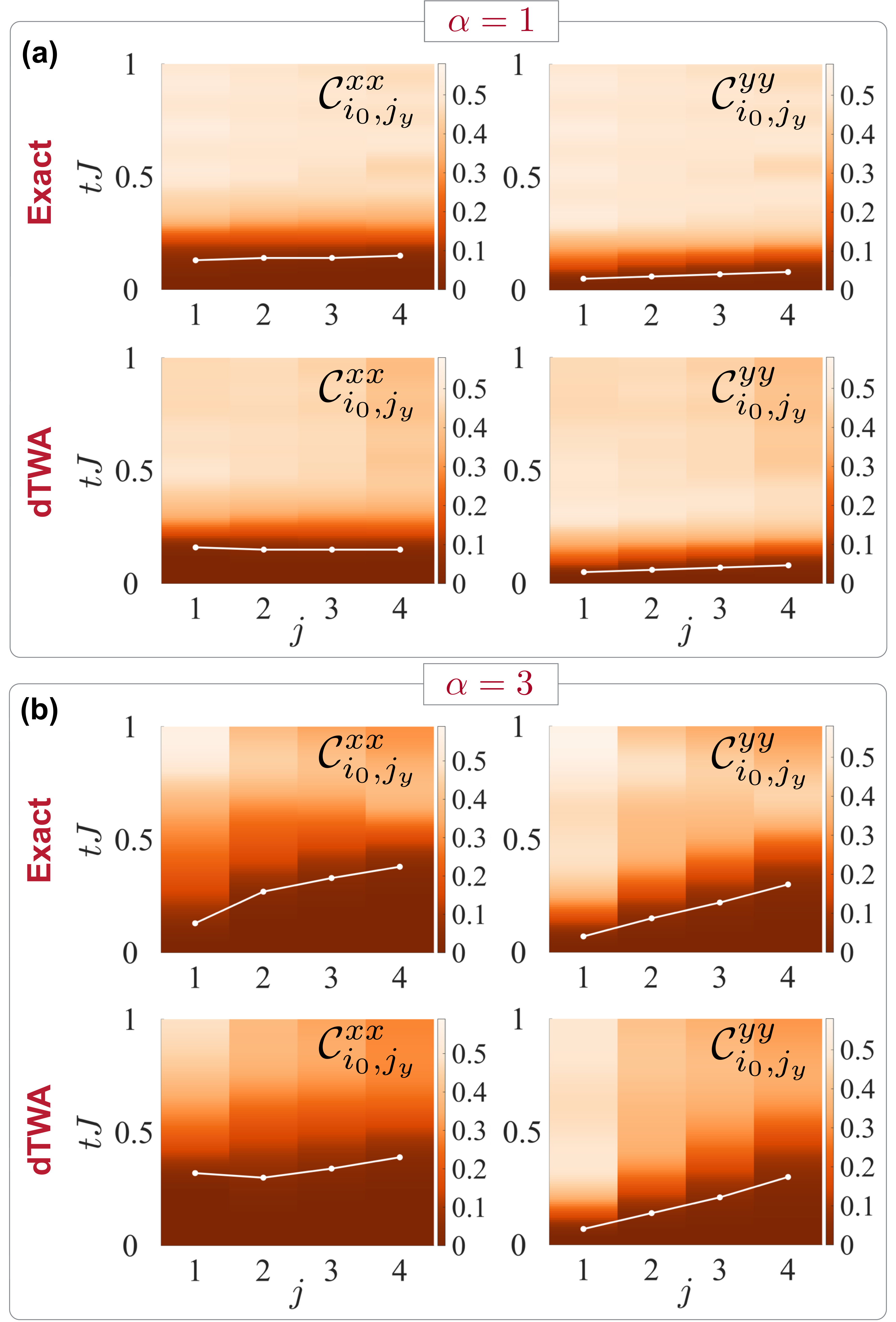}
  \caption{XY interaction benchmark: Time-evolution of connected correlation functions $\mathcal{C}^{xx,yy}_{i_0,j_y}$ in a $4\times5$ lattice between the edge site ${\bf r}_{i_0}=(1,1)$ and the sites ${\bf r}_{j_y}=(1,1+j)$.  Compared are exact diagonalization solutions (upper plots in each panel) and DTWA results (lower plots). (a) $\alpha=1$, (b) $\alpha=3$. The left and right column is for $\mathcal{C}^{xx}_{i_0,j_y}$ and  $\mathcal{C}^{yy}_{i_0,j_y}$, respectively. White lines indicate contours where $\mathcal{C}^{xx,yy}_{i_0,j_y}$ exceeds a threshold value of $\mathcal{C}_{\rm thres}=0.05$. \label{fig:benchmark_corr_xy_al3al1}}
\end{figure}

For the XY model we first compare the DTWA against exact diagonalization results in a small $4\times5$ lattice. Due to the small system size, in order to observe any amount of linear spreading of correlations we have to calculate the correlations from the corner of the system, which leads to additional boundary effects. Specifically, in figure~\ref{fig:benchmark_corr_xy_al3al1} we calculate the correlations between the site with ${\bf r}_{i_0}=(1,1)$ and sites $j$ along the $y$-direction with coordinates ${\bf r}_{j_y}=(1,1+j)$. In this XY case we also  find that the DTWA works (except for deviations at the edge) impressively well, in particular for  $\mathcal{C}^{yy}_{i,j}$ correlations. Most importantly we find that although some oscillation seem not to be well reproduced, the DTWA  accurately captures the spreading of the correlations and the shape of the ``light-cone'' boundary as seen in figure~\ref{fig:benchmark_corr_xy_al3al1}b. The ``light-cone'' boundary at a threshold value $\mathcal{C}_{\rm thres}$ is visualized in  figure \ref{fig:benchmark_corr_xy_al3al1}  by  a contour plot. Physically it  corresponds to  the propagation time required to reach a correlation value of  $\mathcal{C}_{\rm thres}$ between two spins separated by a distance $j$. Here we set  $\mathcal{C}_{\rm thres}=0.05$.

In contrast to the Ising case where the propagation of correlations  barely follows a light-cone spread, in the XY model the situation is more interesting. While for $\alpha=1$, the correlations build up throughout the system almost instantaneously, there is a clear change in behavior when going to shorter range interactions. In the case of $\alpha=3$ a light-cone is expected to emerge \cite{gong_persistence_2014}, and is even visible in the small system calculation shown in figure \ref{fig:benchmark_corr_xy_al3al1}b.  However, in such a small system boundary effects are expected to play an important role.
We point out that the speed of the propagation of correlations in figure~\ref{fig:benchmark_corr_xy_al3al1} is essentially identical for $\mathcal{C}^{xx}_{i,j}$ and $\mathcal{C}^{yy}_{i,j}$. However, the DTWA result for $\mathcal{C}^{xx}_{i,j}$, shows slightly larger discrepancies from the exact solution than $\mathcal{C}^{yy}_{i,j}$. We  also checked the evolution of $\mathcal{C}^{zz}_{i,j}$ (in the XY case), which  exhibits the same type of correlation spreading and is equally well reproduced by the DTWA. However, in our case this particular correlation is  much smaller in magnitude than $\mathcal{C}^{yy}_{i,j}$ and features additional oscillations, which is the motivation to use $\mathcal{C}^{yy}_{i,j}$ in the remainder of this article.

\begin{figure}[tb]
  \centering
    \includegraphics[width=0.8\textwidth]{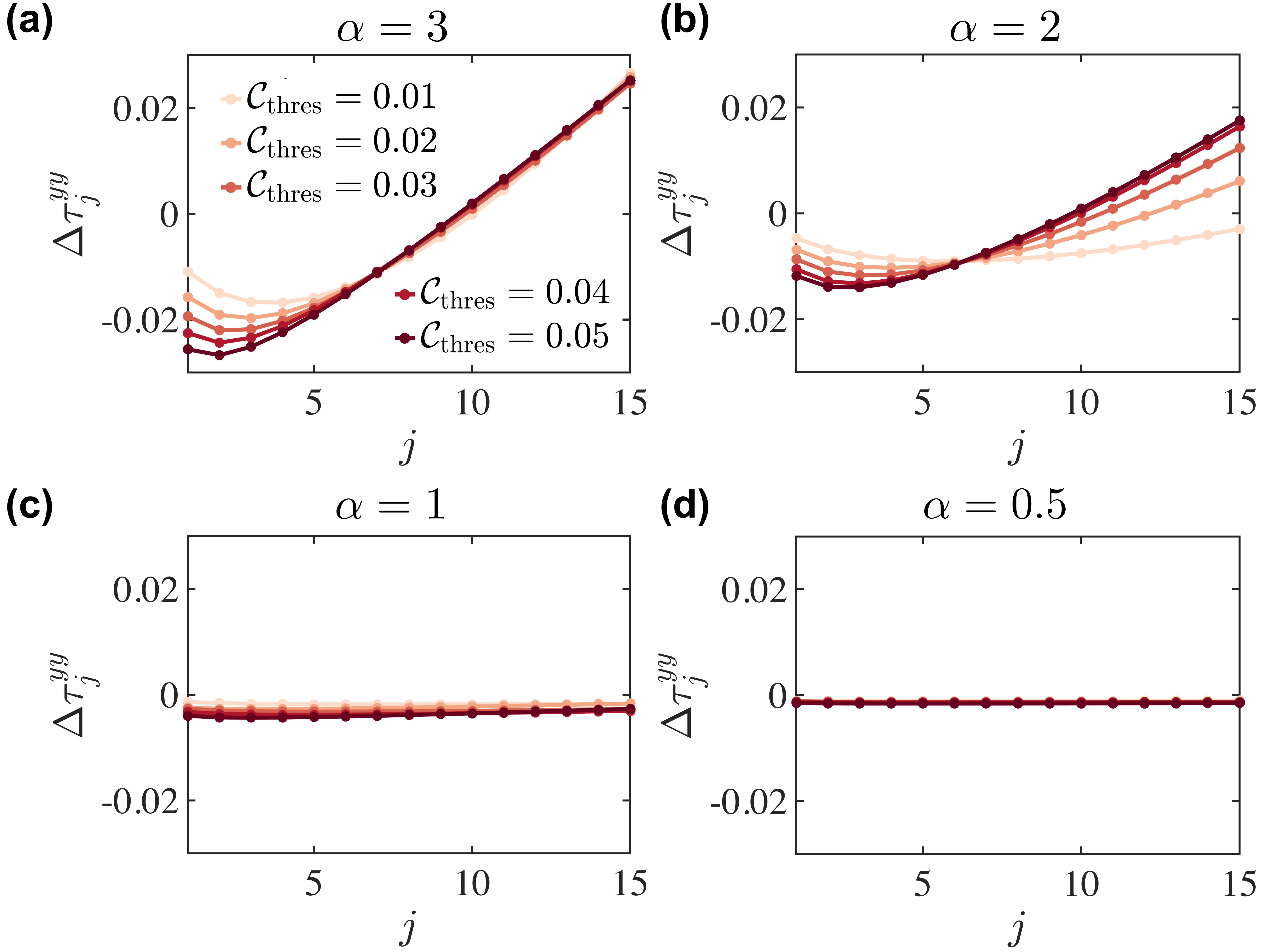}
  \caption{Error analysis. Spatial dependence of the time difference in fitted contour lines. Different colors correspond to different threshold values $\mathcal{C}_{\rm three}=0.01$, $0.02$, $0.03$, $0.04$, $0.05$ (from light to dark color). We compare results in a 1D $1\times31$ lattice for which we can solve the dynamics exactly via t-DMRG methods. Correlations are calculated between sites with ${\bf r}_{i_c}=(1,16)$ and the sites $r_{j_y}=(1,16+j)$. In panels (a)-(d) the interaction range increases, $\alpha=3$, $2$, $1$, $0.5$, respectively. \label{fig:error_analysis}}
\end{figure}

Since  we will be interested in light-cones defined by certain thresholds of correlation values, one has to carefully consider the spatial distribution of the errors as they could lead to wrong conclusions. To test this dependence we compare the results in a 1D system with $1\times 31$ lattice sites, where the correlations are calculated from the center [${\bf r}_{i_c}=(1,16)$]. In this case, exact correlations can be easily computed by means  of t-DMRG techniques \cite{vidal2004, white:real-time_2004,daley:time-dependent_2004,schollwoeck:density-matrix_2011}. To average out statistical noise, we fit a contour to a power law of the form  $\tau_{\mathcal{C}_{\rm thres}}\propto j^\eta$. The error is then calculated as the difference between the DTWA and the exact t-DMRG solution fits via  $\Delta \tau^{yy}_j \equiv  \tau_{\mathcal{C}_{\rm thres}}^{\rm DTWA}-  \tau_{\mathcal{C}_{\rm thres}}^{\rm exact}$. Results for various threshold values and ranges of interactions are shown in figure~\ref{fig:error_analysis}. We find that there are small errors with different sign at different ranges $j$. In case of short-range interactions, for short distance correlations (small $j$), the DTWA predicts a slightly faster growth of correlations, while for long distance correlations (large $j$) it tends to predict a slightly slower growth. For very long-ranged interactions the error becomes very small and homogeneously distributed. Note that in the limit of nearly all-to-all interactions, $\alpha\ll1$, the XY  and the  Ising models become equivalent for fully symmetric initial states. This is because  the XY Hamiltonian becomes a collective spin-Hamiltonian $\propto \hat S_x^2 + \hat S_y^2$, whose dynamics is the same as that of the Ising model due to the conservation of the total collective spin  $\vec S^2$. In this regime the correlations are spatially homogeneous and the error is proportional to $\epsilon_j^{yy}\propto (tJ)^2$.

In conclusion, the behavior of $\Delta \tau^{yy}_j$ with distance will lead to small errors in the predicted power law-exponent $\eta$, which become larger when the range of the interactions becomes shorter. This is also what we observe in figure~\ref{fig:crossover}. In general, there we find that the various values of $\eta$ are still excellently reproduced for a wide range of interaction decay exponents, $\alpha$. This confirms the validity of the  DTWA to capture the propagation of  correlations  in the XY model.

\section{DTWA predictions for spreading of correlations in large systems}

\label{sec:large}

\begin{figure}[b]
  \centering
    \includegraphics[width=0.9\textwidth]{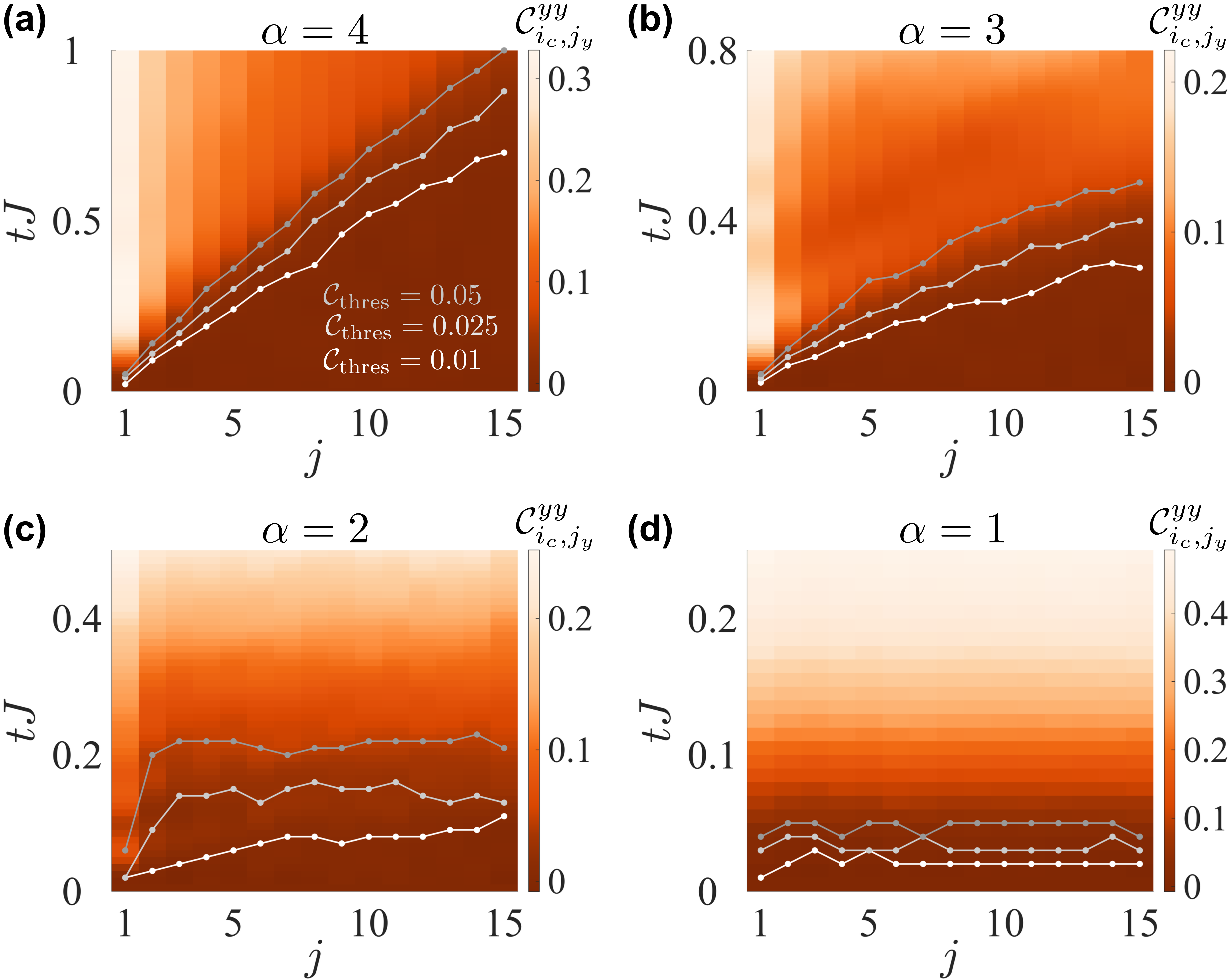}
  \caption{Spreading of  $\mathcal{C}^{yy}_{i_c,j_x}$ correlations in a large $31\times31$ system with XY interactions. (a) $\alpha=4$, (b) $\alpha=3$, (c) $\alpha=2$, and (d) $\alpha=1$. Points where the correlations exceed certain threshold correlations $\mathcal{C}_{\rm thres}=0.01,0.025,0.05$ indicate a light-cone and are shown as contour lines.\label{fig:large_sys}}
\end{figure}

\begin{figure}[b]
  \centering
    \includegraphics[width=0.8\textwidth]{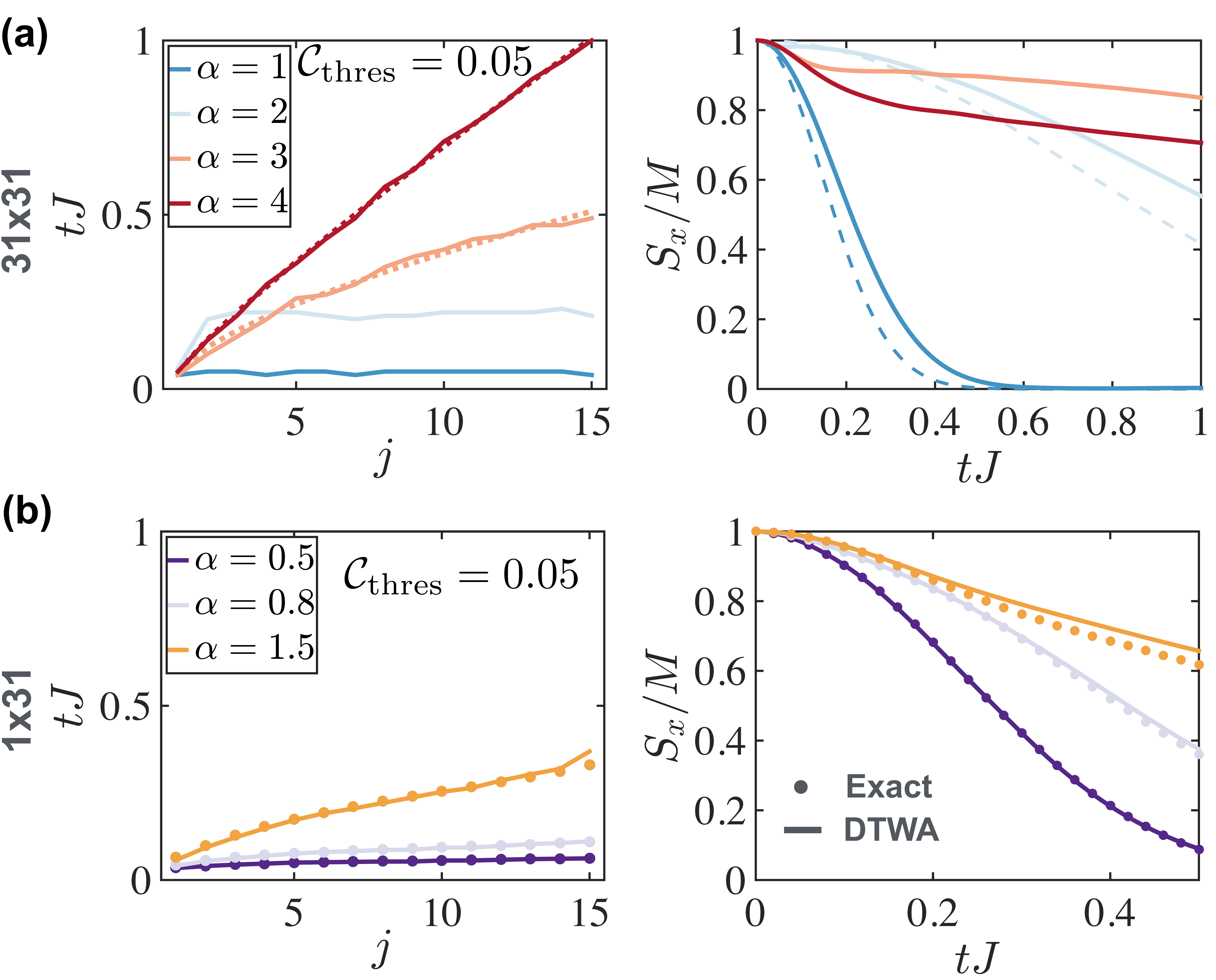}
  \caption{Light cone and contrast dynamics for XY interactions. Left panels: Light-cones indicated by contour lines for a threshold correlation of $\mathcal{C}_{\rm thres}=0.05$. Right panels: decay of  $S_x=\langle \hat S_x \rangle$. (a) A large 2D $31\times31$ system features a rapid transition from light-cone behavior to all-to all physics at $\alpha\sim2$. Also a qualitative change of behavior of the $S_x$ decay occurs at $\alpha \sim 2$.
Left: Power law fits, $\tau_{\mathcal{C}_{\rm thres}}\propto j^\eta$ are shown as dotted lines. Right: Dashed lines show an analytical approximation to $S_x$ (see text).
(b) Similar physics emerges in a 1D $1\times31$ system with a change in behavior of the $S_x$ decay at $\alpha\sim1$. Both exact t-DMRG (points) and DTWA (lines) are shown and consistent.\label{fig:different_alpha}}
\end{figure}

After having validated  the DTWA method, we now use this technique to  calculate dynamics in a large 2D XY model. As in our previous examples for the Ising case, we  focus on a $31\times31$ square lattice geometry and study the evolution of $\mathcal{C}^{yy}_{i_c,j_y}$.
In figure~\ref{fig:large_sys} we show  large system results for decay exponents $\alpha=1,2,3,$ and $4$.  Again, by defining the time where the correlations exceed a certain value $\mathcal{C}_{\rm thres}$, we can define contours that indicate light-cones (see figure~\ref{fig:large_sys} with $\mathcal{C}_{\rm thres}=0.01,0.025$, and $0.05$).  In general  we observe a drastic change in the dynamics of correlations as $\alpha$ is varied.
 While in  the case of short-range interactions, $\alpha=3,4$, we see a clear light-cone-like propagation behavior, for $\alpha=1,2$,  this behavior breaks down rapidly and instead almost instantaneous propagation of correlations is observed. This is summarized in  figure~\ref{fig:different_alpha}a (left panel)  where we show the light-cone boundary for $\mathcal{C}_{\rm thres}=0.05$. In the case of $\alpha=3,4$, we can for example easily fit a power-law curve to the contour of the form $\tau_{{\mathcal C}_{\rm three}} \propto j^\eta$, with a fixed exponent $\eta$.

Interestingly, the rapid change in  propagation behavior is also directly reflected in the time-evolution of the experimentally much more accessible observable $\hat S_x $, as demonstrated in figure~\ref{fig:different_alpha}a.
Linked  to the  disappearance of the light-cone (figure~\ref{fig:different_alpha} left panel) at $\alpha=2$, the behavior of the contrast decay as a function of time (figure~\ref{fig:different_alpha}a, right panel) changes.
For $\alpha=3,4$ after an initial quadratic decay, $\langle \hat S_x \rangle$ decays slowly (remarkably more slowly for $\alpha=3$ than for $\alpha=4$).  For $\alpha \le 2$, however, these two time-scales disappear and  $\langle \hat S_x \rangle$ exhibits a qualitatively different decay. In figure~\ref{fig:different_alpha}b, we show results for the same calculation in a 1D 1$\times31$ lattice, and a corresponding comparison to exact t-DMRG calculations. We observe the same qualitative change in behavior in the decay of $\langle \hat S_x \rangle$, now when crossing $\alpha \sim 1$.

In the 2D case, the behavior for $\alpha\leq 2$ can be understood semi-analytically. For such long-range interactions, as explained above,  we can approximately replace the couplings by a constant: $J_{ij} \approx J_{\rm eff}$, and map the dynamics  to an  Ising Hamiltonian $-J_{\rm eff} \hat S_z^2$. The effective coupling constant $J_{\rm eff}$ in our finite square lattice
($M$ spins in total) can be determined, for example, by requiring that the total energy of the central spin interacting with all other spins with couplings $J_{ij}$ is the same as with coupling $J_{\rm eff}$. From the solution of the Ising model, one thus obtains $ S_x (t) /M=\cos^{M-1}(2 t J_{\rm eff})$ which is shown as dashed lines in figure~\ref{fig:different_alpha}a on the right. Given that there is some arbitrariness in defining a precise value for $J_{\rm eff}$, we see that the contrast decay for long-range interactions is fairly captured by this simple model.

\section{Correlation dynamics crossover with increasing range of interactions}

Although the physics in a 1D chain seems to be relatively similar to the  2D system, a careful examination of figure~\ref{fig:different_alpha} reveals important differences. While in the 2D case, for  $\alpha<2$ the contours are almost  flat, in the 1D case the contours  still seems to exhibit a finite $\eta$ (recall $\tau_{\mathcal{C}_{\rm thres}} \propto j^\eta$ ) even  for $\alpha<1$ (i.e.~for interactions decaying slower than the system's dimensionality). To quantify this observation we systematically evaluate $\eta$ as a function of the range of interactions. Explicitly we vary $\alpha$  between $0.5\lesssim \alpha \lesssim 3$ in 1D and $1.5\lesssim \alpha \lesssim 4$ in 2D and set  the threshold value to be $\mathcal{C}_{\rm thres}=0.05$.

\begin{figure}[tb]
  \centering
    \includegraphics[width=0.8\textwidth]{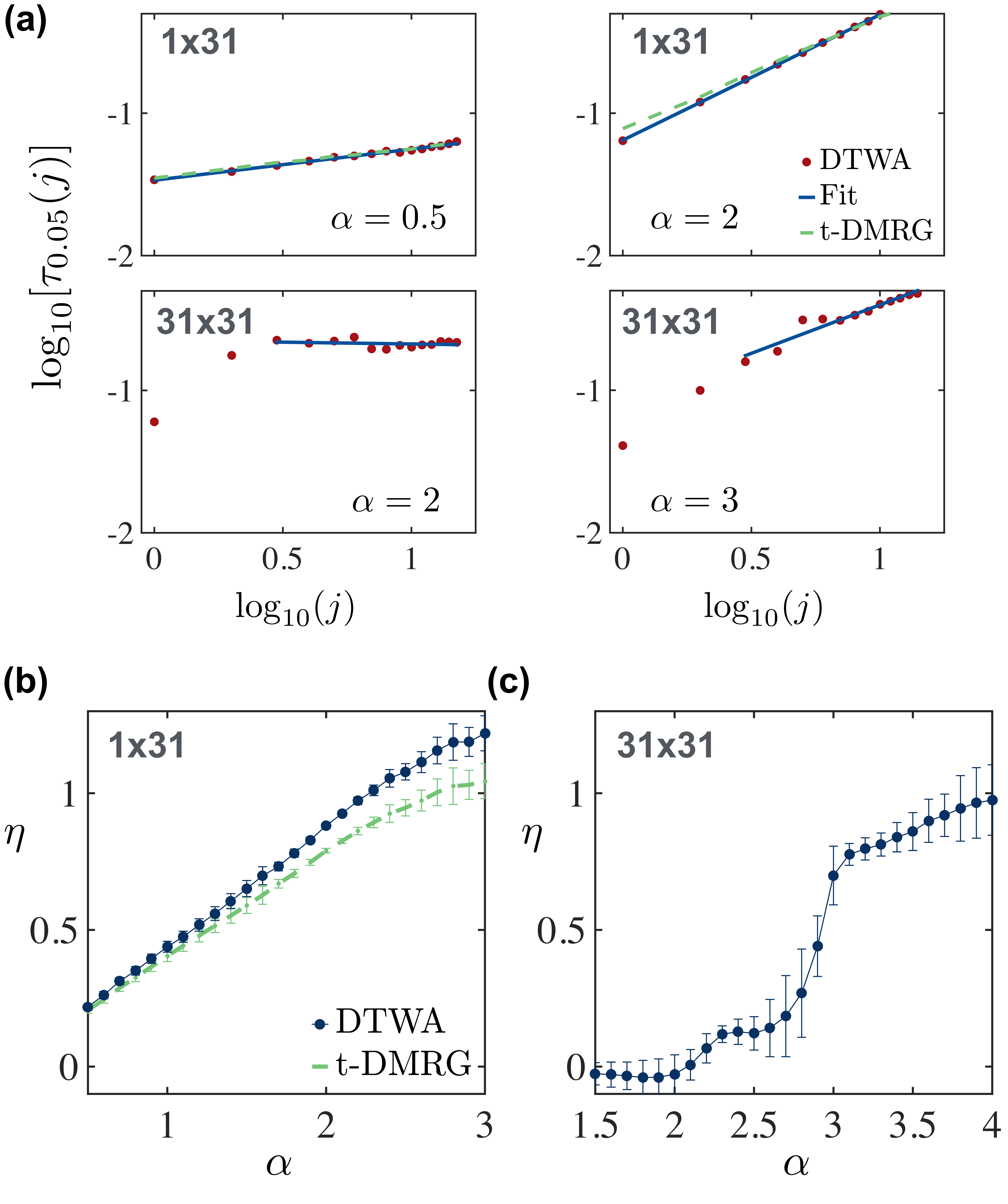}
  \caption{\label{fig:crossover} Correlation dynamics crossover. (a) Plots of the contour lines reveal a  power law behavior, $\tau_{\mathcal{C}_{\rm thres}} \propto \tau^\eta$ (as seen on a log-log scale). Points: DTWA, solid line: power-law fit, dashed line: t-DMRG result. In 2D a clear power law is visible for distances $j\gtrsim 2$, which we use as range for the fit (In 1D we use the whole range).
  The power law exponent, $\eta$, as function of the range of interactions, $\alpha$, is shown in panel (b) for 1D and (c) for 2D.
  }
\end{figure}

In figure~\ref{fig:crossover}a we show selected examples of the light-cone contours for short and long-range interactions in 1D and 2D on a log-log scale (note that in the 1D case we again compare our estimations to exact t-DMRG results). An interesting feature that we observe is that in the 2D case the contour exhibits a clear power law behavior only  for separation  $j\gtrsim2$. We exclude the short distance correlations $j<2$ to perform the linear fit on a log-log scale to avoid this issue.

In 1D (see figure ~\ref{fig:crossover}b)  the DTWA  nicely reproduces the same dependance of $\eta$ vs $\alpha$  seen in the exact t-DMRG calculations (green dashed line). Although it slightly quantitatively over-estimates the light-cone exponent, it shows the correct smooth increase of $\eta$ from zero to $\eta\sim 1$ with increasing $\alpha$. The situation in 2D is strikingly different and instead a rich complicated behavior is observed. Although statistical noise leads to non-negligible error-bars, three clear conclusion can be drawn: i) In contrast to 1D, for $\alpha < 2$ (i.e.~for interactions decaying slower than the system's dimensionality) the power-law exponent of the light cone is consistent with $\eta=0$; ii) There is a sharp increase of $\eta$ at $\alpha=3$; and iii) At $\alpha\sim4$ the light-cone behavior is consistent with a linear causal region ($\eta\sim1$).

Given the  good agreement with exact calculations in the 1D case, we believe that the DTWA predictions are reliable. For the  scenario in consideration (2D XY model with large number of spins), no exact analytical or numerical solution is available, and ultimately experiments need to provide a definite answer.

\label{sec:crossover}

\section{Conclusion \& Outlook}
\label{sec:concl}

We have used a new numerical technique, the DTWA, to study the propagation of correlations in large 2D  XY spin models with long-range interactions, in  regimes accessible to current state-of-the art experiments with polar molecules or trapped ions. We benchmarked this new method in exactly solvable limits (Ising interactions and small systems) and found excellent agreement. In large systems, our method predicts a sharp change in  the dynamics exhibited by  two-point correlation and the Ramsey  contrast when the decay exponent of the interactions $\alpha$ crosses $\alpha=2$. While for $\alpha>2$ a power-law light-cone appears, the DTWA shows an additional jump in the propagation speed of correlations  at $\alpha=3$. For interactions with $\alpha \sim 4$ the DTWA predicts almost linear light-cone behavior.  In the 1D case  a power law light-cone is already seen at decay exponents as low as $\alpha=0.5$, and a nearly linear behavior as  $\alpha\sim3$. We gained confidence in our DTWA prediction by direct  comparisons to exact t-DMRG calculations in 1D.

In the future it will be interesting to study the nature of this sharp transition, not only in 2D, but also in 3D. This is a regime  currently accessible with  polar molecule experiments that encode the spin degree of freedom in  rotational states coupled by  dipolar interactions. In such setups sharp changes in the speed of correlation propagations could be observable. In this implementation  it will be intriguing to investigate the role  played by the anisotropic character of the interactions and  the finite filling fraction on the light-cone dynamics. Systems where retardation effects in the dipolar interaction become relevant (e.g.\ with atoms in two   electronic states \cite{Olmos2013,Chang2004}) could also  become  excellent laboratories for the observation of  DTWA predictions. In many implementations of spin models, dissipation effects (due to for example spontaneous emission or cooperative radiation) compete with the   pure Hamiltonian dynamics. In order to model these experimentally relevant situations  it   will be important to adapt our technique to a master equation formulation instead of  pure Hamiltonian evolution.

\section*{Acknowledgements}
 This
work has been financially supported by  JILA-NSF-PFC-1125844,
NSF-PIF-1211914, ARO,  AFOSR, AFOSR-MURI. Computations utilized the Janus supercomputer, supported by NSF (award number CNS-0821794), NCAR, and CU Boulder/Denver.

\pagebreak

\bibliographystyle{iopart-num}
\bibliography{./molref5d}

\end{document}